\documentclass{elsart}
\usepackage{graphicx}
\usepackage{color}
\usepackage{bm}
\usepackage{multirow}
\usepackage{epsfig}
\usepackage[english]{babel}

\begin{document}
\bibliographystyle{try}
\topmargin 0.1cm
\newcommand{\bc}           {\begin{center}}
\newcommand{\ec}           {\end{center}}
\newcommand{\bq}           {\begin{eqnarray}}
\newcommand{\eq}           {\end{eqnarray}}
\newcommand{\be}           {\begin{equation}}
\newcommand{\ee}           {\end{equation}}
\newcommand{\bi}           {\begin{itemize}}
\newcommand{\ei}           {\end{itemize}}

\newcounter{univ_counter}
\setcounter{univ_counter} {0} \addtocounter{univ_counter} {1}
\edef\HISKP{$^{\arabic{univ_counter}}$ }
\addtocounter{univ_counter}{1}
\edef\GATCHINA{$^{\arabic{univ_counter}}$ }

\begin{frontmatter}

\title{Evidence for a spin-quartet of nucleon resonances at 2\,GeV
}

\author[HISKP,GATCHINA]{A.V.~Anisovich},
\author[HISKP]{E.~Klempt},
\author[HISKP,GATCHINA]{V.A.~Nikonov},
\author[HISKP,GATCHINA]{A.V.~Sarantsev},
\author[HISKP]{U.~Thoma}
\\

\address[HISKP]{Helmholtz-Institut f\"ur Strahlen- und Kernphysik der
Universit\"at Bonn, Germany}
\address[GATCHINA]{Petersburg
Nuclear Physics Institute, Gatchina, Russia}

\date{\today}


\begin{abstract}
Results from a multi-channel partial wave analysis of elastic and
inelastic $\pi N$ and $\gamma N$ induced reactions are presented.
The analysis evidences the existence of a spin-quartet of nucleon
resonances with total angular momenta $J^P=1/2^+,\cdots, 7/2^+$. All
states fall into a $\pm130$\,MeV mass gap centered at 1.97\,GeV. The
spin quartet is at variance with S-wave diquark configurations
required in classical di-quark models.
\vspace{2mm}   
\end{abstract}

\vskip 5mm

\end{frontmatter}

The aim to improve our understanding of the confinement mechanism
and of the dynamics of quarks and gluons in the non-perturbative
region of QCD is the driving force behind the intense efforts to
clarify the spectrum of meson \cite{Klempt:2007cp} and baryon
\cite{Klempt:2009pi} resonances. The systematics of the baryon
ground states were constitutive for the development of quark models.
In the harmonic oscillator (h.o.) approximation, the quark model
predicts a ladder of baryon resonances with equidistant squared
masses, alternating with positive and negative parity, and this
pattern survives approximately in more realistic potentials (see,
e.g. \cite{Capstick:1986bm,Loring:2001kx}). Recent lattice gauge
calculations \cite{Edwards:2011jj} confirm these findings. However,
masses of resonances with positive and negative parities are often
similar, in striking disagreement with quark models and the results
on the lattice. A second problem of both, lattice calculations and
quark models, is the number of expected states which is considerably
larger than confirmed experimentally, a fact which is known as
problem of {\it missing resonances.} The number of expected states
is much reduced if it is assumed that two quarks form a quasi-stable
diquark \cite{Anselmino:1992vg}. Of particular importance is thus
the possible existence of a quartet of positive-parity nucleon
resonances with total angular momentum $J=1/2,\cdots, 7/2$ -- called
$N_{1/2^+}$, $N_{3/2^+}$, $N_{5/2^+}$, $N_{7/2^+}$ here and expected
at about 2\,GeV -- with wave functions which in h.o. approximation
have intrinsic angular orbital momentum $L=2$ and a total quark-spin
angular momentum $S=3/2$. In quark models, the three quarks are
antisymmetric with respect to their color degree of freedom, the
states have a symmetric spin wave function, the nucleon flavor wave
function is of mixed-symmetry. The spatial wave function must hence
be of mixed symmetry, too. This is impossible when one quark pair
forms a diquark in $S$-wave. The existence of these four states
would rule out any diquark model respecting the Pauli principle. It
is this consequence that led Nathan Isgur \cite{Isgur:1988pr}, more
than 20 years ago, to consider the existence of this quartet to be
one of the most important open issues in baryon spectroscopy.
Unfortunately, this question is still not yet answered.

Experimentally, a few candidates for positive-parity nucleons in the
1.95\,GeV region have been reported. The Particle Data Group
\cite{Nakamura:2010zzi} lists three states - $N_{7/2^+}(1990)$,
$N_{5/2^+}(2000)$, $N_{3/2^+}(1900)$ - the evidence for their
existence is estimated to be fair. Only one of them,
$N_{7/2^+}(1990)$, was reported by both classical analyses at
Karlsruhe-Helsinki (KH) \cite{Hohler:1979yr} and Carnegie Mellon
(CM) \cite{Cutkosky:1980rh} exploiting elastic $\pi$-nucleon
scattering, $N_{5/2^+}(2000)$ is seen by \cite{Hohler:1979yr} only,
none is seen in the more recent analysis at George Washington
University (GW) \cite{Arndt:2006bf} which include high-precision
data from meson factories and measurements of the phase-sensitive
polarization state of the recoiling nucleon. Obviously, it is
extremely hard to decide on the existence of these states on the
basis of elastic $\pi N$ scattering: the coupling of these states to
$\pi N$ is too weak to leave a recognizable trace in the data.

In this letter we present evidence for the full spin quartet of
positive parity nucleon resonances in the 1.88 to 2.02\,GeV mass
range from a partial wave analysis of a large variety of pion- and
photo-induced reactions. The list of all reactions used in this
analysis, more details on the partial wave amplitudes, and a full
account of the results can be found elsewhere
\cite{Anisovich:2010an,Anisovich:2011ye}. Of decisive importance for
the results reported here are the recent high-precision data on
photoproduction of the $\Lambda$ hyperon. In the reaction $\gamma
p\to \Lambda K^+$, the initial photon can be polarized linearly or
circularly, the polarization of the outgoing $\Lambda$ can be
constructed from its parity-violating $\Lambda\to p\pi^-$ decay.
Thus not only  the differential cross sections $d\sigma/d\Omega$ but
also the induced $\Lambda$ polarization $P$, the beam asymmetry
$\Sigma$   and spin-correlation coefficients, called $C_x, C_z$,
$O_x, O_z$, can be determined
\cite{Lleres:2007tx,Lleres:2008em,Castelijns:2007qt,McCracken:2009ra}.
To construct the multipoles governing the process for discrete
energy bins in an energy-independent partial wave analysis,
experiments with polarized target protons would be required in
addition. Such data have been taken but are not yet published. On
the other hand, the experimental observation that the squared sum of
a reduced set of polarization observables $P^2+C_x^2+C_z^2$ -- which
is bound to be less than one \cite{Artru:2006xf} -- actually
exhausts unitarity \cite{Bradford:2006ba} may serve as a hint that
the available data may be sufficient to arrive at a nearly
model-independent partial wave solution.

\begin{table} \caption{\label{res-list}List of nucleon
resonances used in the coupled channel analysis. The Breit-Wigner
masses and widths (in MeV) are given as small numbers.}
\bc
\renewcommand{\arraystretch}{1.3}
\begin{tabular}{cccccc} \hline\hline
$N_{1/2^-}(1535)$&\hspace{-3mm}$N_{1/2^-}(1650)$&\hspace{-3mm}$N_{1/2^-}(1895)$&\hspace{-3mm}$N_{3/2^-}(1520)$
&\hspace{-3mm}$N_{3/2^-}(1700)$&\hspace{-3mm}$N_{3/2^-}(1875)$\\[-1.4ex]
 \tiny 1519$\pm$5; 128$\pm$14&\hspace{-3mm} \tiny 1651$\pm$6; 104$\pm$10&\hspace{-3mm} \tiny 1895$\pm$15;
$ 90^{+30}_{-15}$
 &\hspace{-3mm} \tiny 1517$\pm 3$; 114$\pm$5&\hspace{-3mm} \tiny 1790$\pm$40;
 390$\pm$140
 &\hspace{-3mm} \tiny 1880$\pm$20; 200$\pm$25\\[-0.6ex]
$N_{3/2^-}(2150)$&\hspace{-3mm}$N_{5/2^-}(1675)$&\hspace{-3mm}$N_{5/2^-}(2060)$&\hspace{-3mm}$N_{7/2^-}(2190)$&\hspace{-3mm}$N_{9/2^-}(2250)$&\hspace{-3mm}\\[-1.4ex]
 \tiny 2150$\pm$60; 330$\pm$45&\hspace{-3mm} \tiny 1664$\pm$5; 152$\pm$7&\hspace{-3mm} \tiny 2060$\pm$15;
 375$\pm$25
 &\hspace{-3mm} \tiny 2180$\pm$20; 335$\pm$40&\hspace{-3mm}\tiny 2280$\pm$40;
 520$\pm$50\\
$N_{1/2^+}(1440)$&\hspace{-3mm}$N_{1/2^+}(1710)$&\hspace{-3mm}$N_{1/2^+}(1880)$
&\hspace{-3mm}$N_{3/2^+}(1720)$&\hspace{-3mm}$N_{3/2^+}(1900)$
&\hspace{-3mm}$N_{5/2^+}(1680)$\\[-1.4ex]
 \tiny 1430$\pm$8; 365$\pm$35&\hspace{-3mm} \tiny 1710$\pm$20; 200$\pm$18&\hspace{-3mm} \tiny 1870$\pm$35;
 235$\pm$65
 &\hspace{-3mm} \tiny 1690$^{+70}_{-35}$; 420$\pm$100
 &\hspace{-3mm} \tiny 1905$\pm$30; 250$^{+120}_{-\ 50}$&\hspace{-3mm} \tiny 1689$\pm$6; 118$\pm$6\\[-0.6ex]
$N_{5/2^+}(1860)$&\hspace{-3mm}$N_{5/2^+}(2000)$&\hspace{-3mm}$N_{7/2^+}(1990)$&\hspace{-3mm}$N_{9/2^+}(2200)$
&\hspace{-3mm}&\hspace{-3mm}\\[-1.4ex]
 \tiny 1860$^{+120}_{\ -60}$;
250$^{+150}_{\ -50}$&\hspace{-3mm}  \tiny 2090$\pm$120;
460$\pm$100&\hspace{-3mm} \tiny 2050$\pm$60; 210$\pm$70
&\hspace{-3mm} \tiny 2200$\pm$50; 480$\pm$60\\
\hline\hline
\end{tabular}\renewcommand{\arraystretch}{1.0}\ec
\end{table}

The partial wave amplitude $A^{(\beta)}_{ij}$ is given by a K-matrix
which incorporates a summation of resonant and non-resonant terms in
the form
\be
A^{(\beta)}_{ij}=\sqrt{\rho_i} \sum\limits_a
K^{(\beta)}_{ia}\left(I-i\rho K^{(\beta)}\right
)^{-1}_{aj}\sqrt{\rho_j}\,\,.
\label{res_amp}
\ee
It describes transitions e.g. from the initial state $i=N\pi$ to the
final state $j=\Lambda K^+$. The multi-index $\beta$ denotes the
quantum numbers of the partial wave, the factor $\rho$ represents
the phase space matrix to all allowed intermediate states,
$\rho_{i}$, $\rho_{j}$ the phase space of the initial and the final
state. An introduction to the K-matrix approach in hadron
spectroscopy can be found elsewhere \cite{Chung:1995dx}. Resonances
are represented by K-matrix poles decaying into channels with
largest cross sections. In most waves the K-matrix are coupled to
the channels (with indices $i,a,j$) $N\pi$, $N\eta$, $\Lambda K$,
$\Sigma K$, $\Delta\pi$, and $N m$ where the latter describes
missing channels (mainly $N\rho /\omega$). Weak channels like
$N_{1/2^+}(1440)\pi$, $N(\pi^0\pi^0)_{\rm S-wave}$,
$N_{3/2^-}(1520)\pi$, $N_{1/2^-}(1535)\pi$, $N_{5/2^+}(1675)\pi$ are
fitted in the framework of the D-vector approach
\cite{Anisovich:2007bq}. The decay of resonances into these channels
are taken into account only at the last interaction vertex. However,
if we find that the coupling to a particular channel is large
enough, this channel is included explicitly to the K-matrix
parameterization and data are refitted. For example, the nucleon
$J^P=3/2^+$ K-matrix includes in addition $N(\pi^0\pi^0)_{\rm
S-wave}$ and $N(1520)\pi$ channels. The K-matrix poles are real
numbers, the complex pole positions are given by the poles of the
partial wave amplitudes $A^{(\beta)}_{ij}$ in the complex energy
plane. Table \ref{res-list} lists the nucleon resonances used in the
fits. Background K-matrix terms are derived from meson ($\pi, \rho,
K, K^*$) exchanges in the $t$-channel and nucleon, $\Delta$,
$\Lambda$, and $\Sigma$ exchange in the crossed ($u$-) channel.

\begin{figure*}[h]
\bc
\begin{tabular}{cc}
\epsfig{file=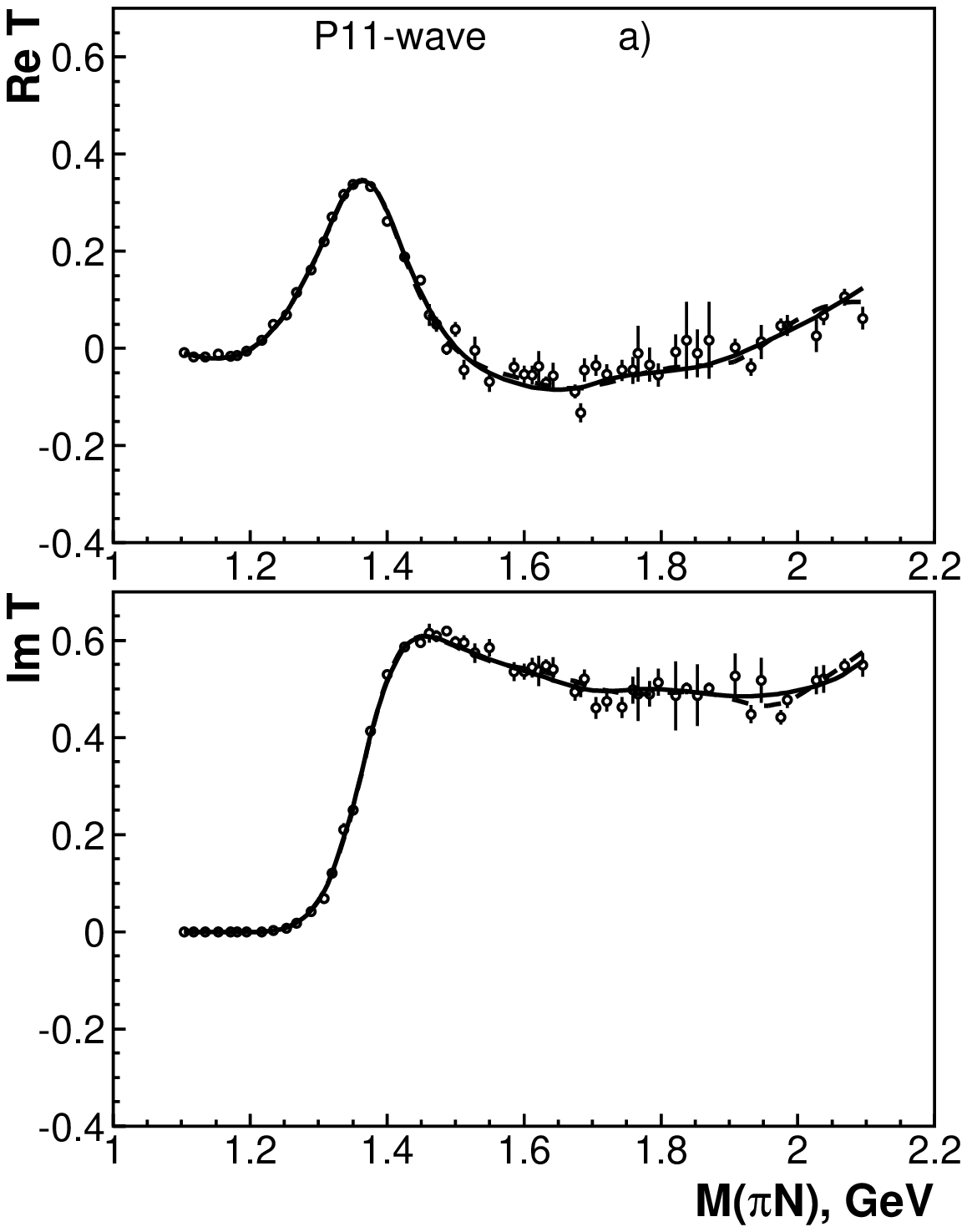,width=0.40\textwidth,clip=on}&
\epsfig{file=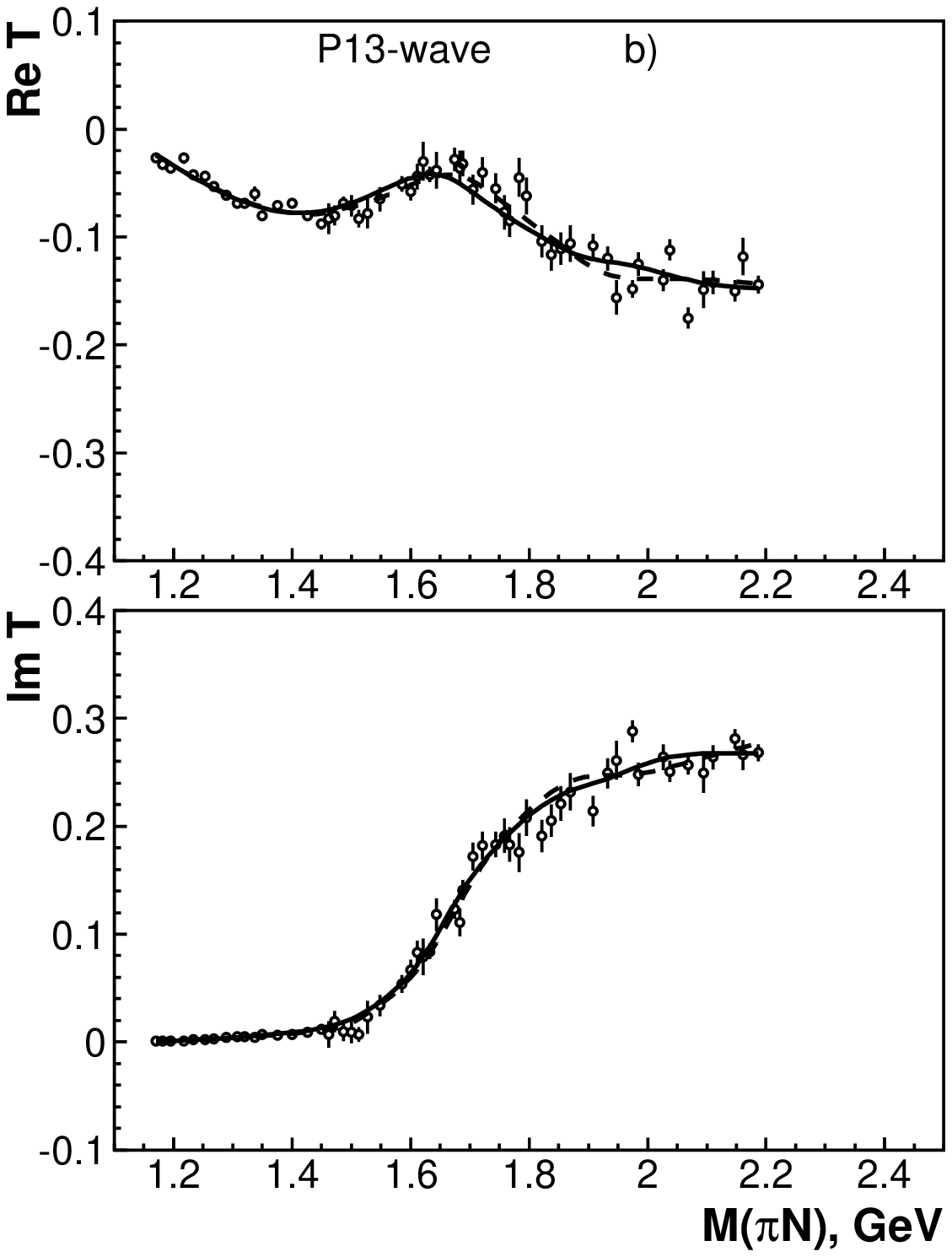,width=0.40\textwidth,clip=on}\\
\epsfig{file=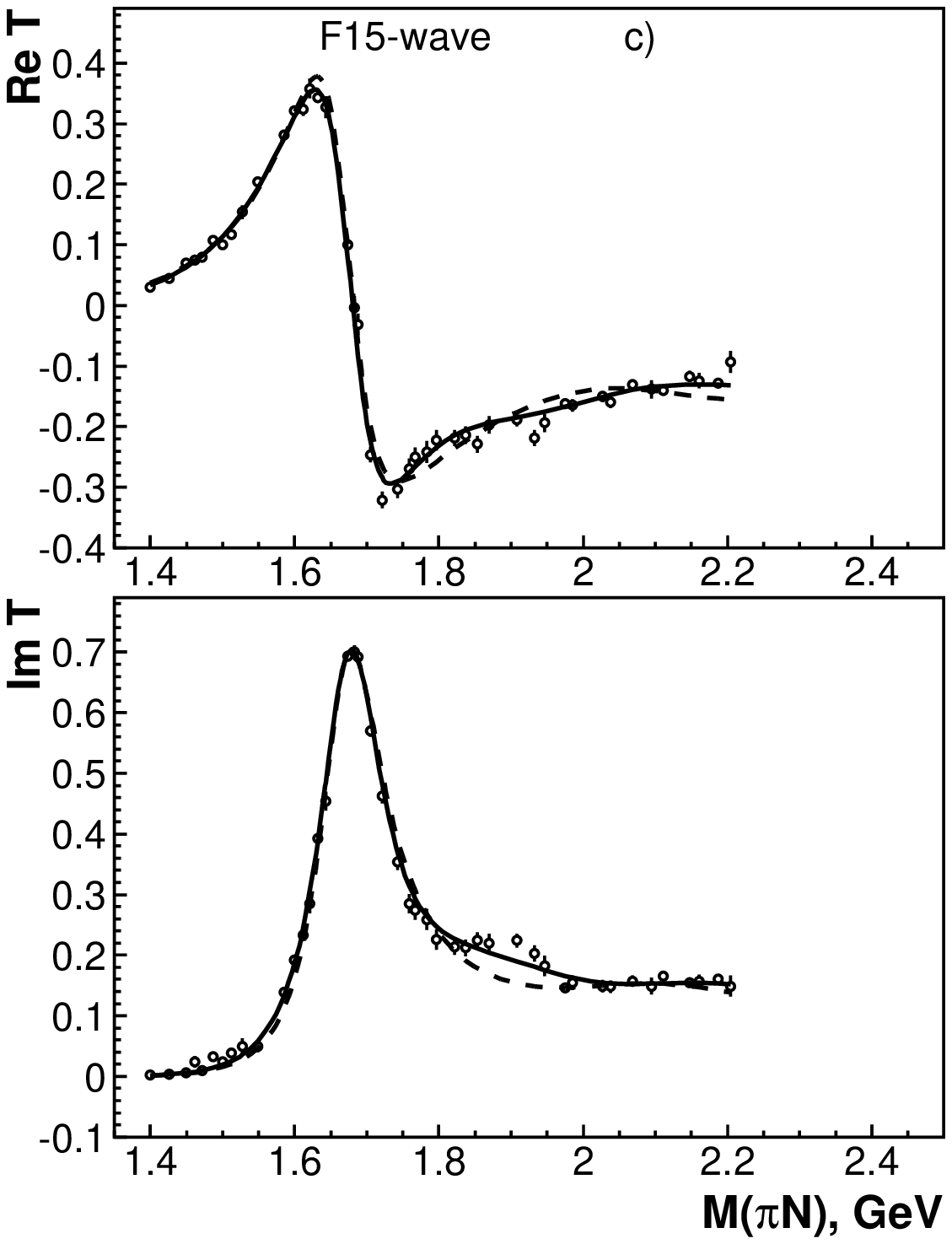,width=0.40\textwidth,clip=on}&
\epsfig{file=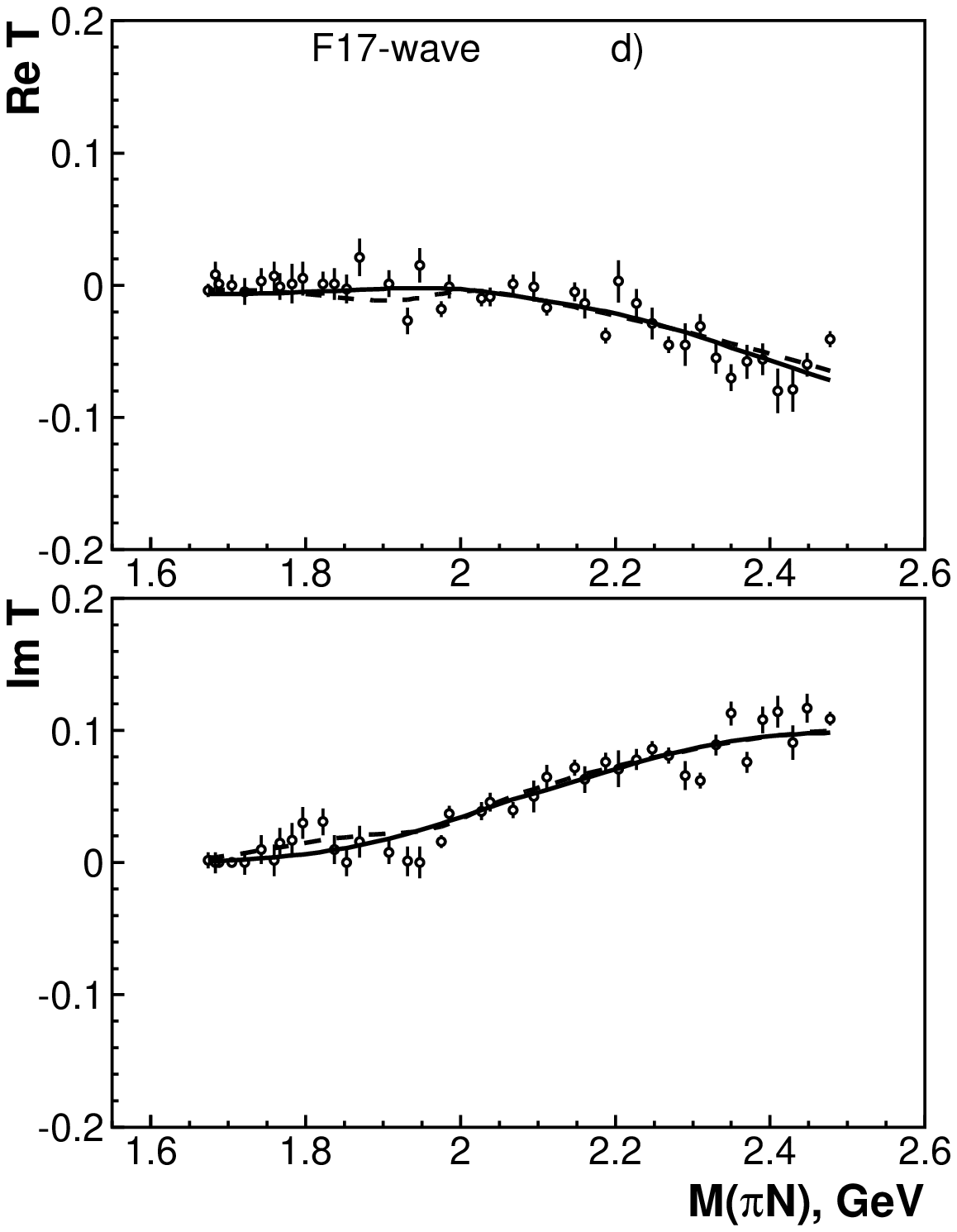,width=0.40\textwidth,clip=on}\\
\end{tabular}
\ec
\caption{\label{elastic}Elastic partial wave amplitudes (real and
imaginary parts) for $J^P=1/2^+$ (a), $J^P=3/2^+$ (b), $J^P=5/2^+$
(c) and $J^P=7/2^+$ (d). The points represent the energy independent
partial-wave amplitudes from \cite{Arndt:2006bf}. For the partial
waves $J^P=1/2^+$, $J^P=3/2^+$ and $J^P=7/2^+$ the dashed curves
correspond to the solution BG2011-01 and solid curves to the
solution BG2011-02. For $J^P=5/2^+$ the solid curve corresponds to
the three pole and dashed to the two pole K-matrix solutions.}
\end{figure*}

Fig. \ref{elastic} shows elastic $\pi N$ positive-parity partial
wave amplitudes for nucleons with $J=1/2^+, \cdots, 7/2^+$. The
energy-independent amplitudes \cite{Arndt:2006bf} are compared with
the result from our coupled-channel fits. It is at this point that
the recent achievements in data and partial wave analysis methods
enter. Fits to the elastic $\pi N$ scattering data experience
inelastic reactions like $\pi^-p\to\Lambda K^+$ as loss in the $\pi
N\to\pi N$ partial wave. But from $\pi N$ elastic scattering alone,
there is no knowledge about the dynamics of the inelastic reactions.
By use of the full existing data base, this gap is filled. The
reactions $\gamma p\to p\pi^0$ and $\gamma p\to n\pi^+$ provide a
bridge from pion-induced reactions to the rich field of
photoproduction. Inelastic reactions are no longer a black box, they
are constrained by high-precision data. Yet, many data do not cover
the full angular range, and polarization information is still not
complete. Presumably, this is the reason why the fit to the data
does not converge to a well-defined minimum. However, in minima with
a good $\chi^2$, the properties of baryon are rather stable. Their
spread in a large variety of fits is used to define error bars. In
some partial waves, the results are ambiguous; then both solutions
are discussed. The two solutions are called BnGa2011-1 and
BnGa2011-2. For technical reasons, reactions with three-body final
states are not included in the fits in mass scans described below.

We now turn to a discussion of individual partial waves with
$J^P\!=\!\frac12^+$,$\frac32^+$,$\frac52^+$,$\frac72^+$.

{\boldmath $I(J^P)=\frac12(\frac12 ^+)$:\qquad} The contribution of
the $J^P=\frac12^+$ wave in the $\pi^-p\to K^0\Lambda$ reaction
maximizes at 1720\,MeV, followed by a long tail.  Thus a significant
contribution from $N_{1/2^+}(1710)$ should be expected. The long
tail may indicate the need for a further resonance at higher mass.
We therefore used a four-pole K-matrix amplitude to fit the data: a
first pole at the nucleon mass as Born term, followed by the well
known Roper resonance at 1440\,MeV \cite{:2007bk}, and two further
poles. If both these resonances are omitted from the fit, the fit
exhibits problems to reproduce the differential cross sections and
the recoil asymmetry. The contributions from the $J^P=\frac12^+$
wave to $\pi^-p\to K^0\Lambda$ and $\pi^-p\to K^0\Sigma$ become
smaller and featureless. The description significantly improves when
$N_{1/2^+}(1710)$ is introduced. Its pole is found at $(1687\pm 17)
-i\,(95\pm 17)$\,MeV, averaged over all solutions BG2011. The
overall description is notably improved when a further K-matrix pole
in the 1870\,MeV region is introduced in the fit. The resonance was
first suggested in \cite{Castelijns:2007qt}. The pole position of
the amplitude optimizes for  $M_{\rm pole}=1905\!\pm\!35-
i(85\!\pm\!20)$\,MeV in the first class of the solutions and at
$M_{\rm pole}=1870\!\pm\!25-i(140\!\pm\!18$)\,MeV in the second
class of the solutions. We call this resonance $N_{1/2^+}(1880)$;
the Breit-Wigner mass which corresponds to the pole is at $1870\pm
35$\,MeV.

\begin{figure*}[pt]
 \bc
\epsfig{file=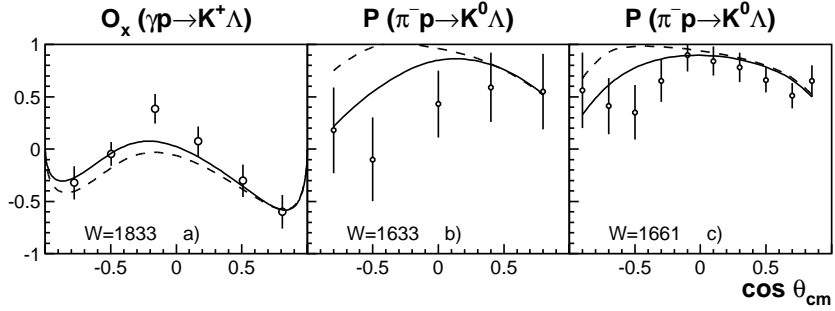,width=0.8\textwidth,clip=on}
 \ec
\caption{\label{select}Selected distributions with fits including
(solid line) and excluding (dashed line) $N_{1/2^+}(1880)$ from the
list of resonances included in the fits. The data (a) are from
\cite{Lleres:2008em} and (b,c) from
\cite{Baker:1978qm,Saxon:1979xu}.}
\end{figure*}

We did not find a comparable solution when the amplitude was
parameterized as a three-pole K-matrix plus the Breit-Wigner
amplitude for $N_{1/2^+}(1880)$. This type of fits produced $\chi^2$
values which were much worse than the one obtained in the four-pole
K-matrix solution. Nevertheless, the best mass of the Breit-Wigner
state was found at 1855\,MeV, but the contribution of this state to
the cross sections is very small. It had very weak pion-nucleon and
photo-couplings. As the result, a mass scan - as done for the other
resonances discussed below - did not show a clear minimum. Thus we
conclude that this state creates an interference pattern which can
be reproduced only in an approach which takes into account the
interference between different poles properly. The situation is
similar to the interference of the two $J^P=\frac12 ^-$ resonances
at 1535 and 1650\,MeV. Their pattern can also not be described with
reasonably accuracy using Breit-Wigner amplitudes. A scan like the
ones shown for the $N_{3/2^+}(1900)$, $N_{5/2^+}(2000)$, and
$N_{7/2^+}(1990)$ is thus not possible. Instead, we show in
Fig.~\ref{select} some selected data and a fit with
$N_{1/2^+}(1880)$ and a fit when $N_{1/2^+}(1880)$ is removed from
the list of resonances. The overall $\chi^2$ improvement is 2090 for
the reactions with two body final states when $N_{1/2^+}(1880)$ is
introduced.

{\boldmath $I(J^P)=\frac12(\frac32 ^+)$:\qquad} Evidence for a
$N(1900)P_{13}$ resonance was reported by Manley and Saleski
\cite{Manley:1992yb}, Penner and Mosel \cite{Penner:2002ma}, Nikonov
{\it et al.} \cite{Nikonov:2007br}, and Schumacher and
Sargsian~\cite{Schumacher:2010qx}. In the analysis of Nikonov {\it
et al.}, its existence followed from a multichannel fit which
included the double polarization observables $C_x$, $C_z$
\cite{Bradford:2006ba} in photoproduction of kaon-hyperon final
states. The solution predicted very well the data on further double
polarization observables, $O_x$, $O_z$, reported one year later
\cite{Lleres:2008em}. The new data on $\gamma p\to \Lambda K^+$
strengthen the need for this resonance. In a fit with a two-pole
K-matrix, the pole position is determined to $(1915\pm 50) -
i\,(90\pm 25)$\,MeV. The fit is improved when a three-pole structure
is assumed. In solution BG2011-02, the two high-mass poles are found
at $(1870\!\pm\! 35) - i\,(155\!\pm\!20)$\,MeV and $(1955\pm 30) -
i(115\pm 20)$\,MeV, respectively. The latter resonance is
tentatively called $N_{3/2^+}(1975)$ here. In the solution
BG2011-01, this structure is not observed. Here the second pole is
seen at $(1910\pm 12)-i(100\pm 8)$\,MeV and the third pole is ill
defined only. It has a rather large width (more than 600\,MeV) and
its mass is located in the region 2100-2300\,MeV.

We should expect that two close-by resonances like $N_{3/2^+}(1900)$
and $N_{3/2^+}(1975)$ should chose different decay modes. A well
known example for such a pattern are the decays of the two meson
resonances $K_1(1280)$ and $K_1(1400)$ where the first resonance
decays prominently into $K\rho$, the latter one into $K^*\pi$. Here,
only the photo-couplings of $N_{3/2^+}(1900)$ and $N_{3/2^+}(1975)$
are distinctively different and, perhaps, the branching ratio for
$N\eta$ decay; the other branching ratios are similar in magnitude.
Yet, the branching ratios observed here are rather small. Large
differences might show up, e.g., in the branching ratios for decays
into $\Delta\pi$ and into $N\rho$ or other multibody decay modes.
Certainly, the existence of two resonances, $N_{3/2^+}(1900)$ and
$N_{3/2^+}(1975)$, cannot be considered as established. The
existence of $N_{3/2^+}(1900)$ is, however, mandatory to achieve an
acceptable fit in both classes of the solutions.

\begin{table}
\caption{\label{BR}The decay branching ratios of
$N_{3/2^+}(1900)P_{13}$ and $N_{3/2^+}(1975)$ when the existence of
two resonances is assumed. See text for the discussion.\vspace{3mm}}
\bc
\begin{tabular}{ccc} \hline\hline
 Channel&$N_{3/2^+}(1900)$&$N_{3/2^+}(1975)$\\
 \hline
 $Br(\pi N)$           & $4\pm 2 \%$     & $1.5\pm 1 \%$    \\
 $Br(\eta N)$          & $11  \pm 5 \%$   & $2\pm 1 \%$    \\
 $Br(K\Lambda)$        & $13 \pm 5 \%$     & $6\pm 3 \%$    \\
 $Br(K\Sigma)$         & $6\pm 3 \%$       & $5\pm 2 \%$    \\
 \hline
$A_{1/2}$  {\scriptsize $10^{-3}$GeV$^{-1/2}$} &$25\!\pm\!17$\,/$40\!\pm\!30^o$ &-$17\!\pm\!8$\,/$-70\!\pm\!30^o$ \\
$A_{3/2}$  {\scriptsize $10^{-3}$GeV$^{-1/2}$}&$95\!\pm\!20$\,/$110\!\pm\!30^o$ &$65\!\pm\!20$\,/-$70\!\pm\!30^o$\\
\hline\hline
 \end{tabular}
\ec
 \end{table}

\begin{figure*}[pt]
 \bc
\begin{tabular}{cccc}
\hspace{-3mm}\epsfig{file=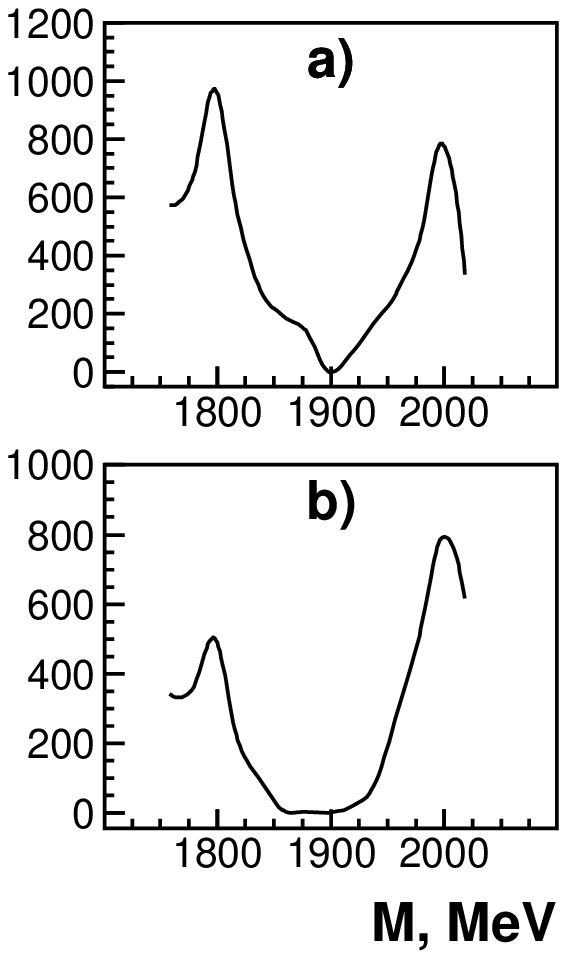,width=0.32\textwidth,clip=on}&
\hspace{-3mm}\epsfig{file=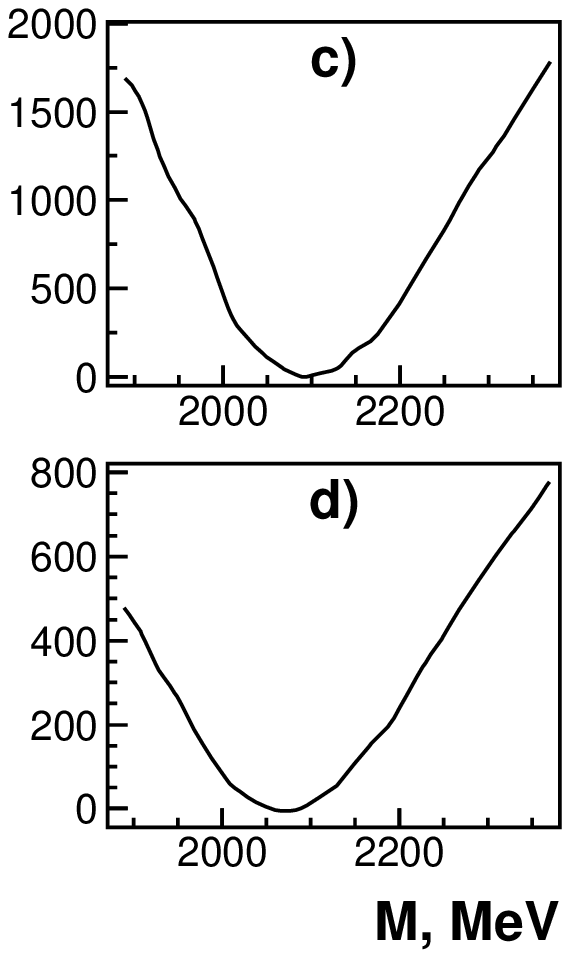,width=0.32\textwidth,clip=on}&
\hspace{-3mm}\epsfig{file=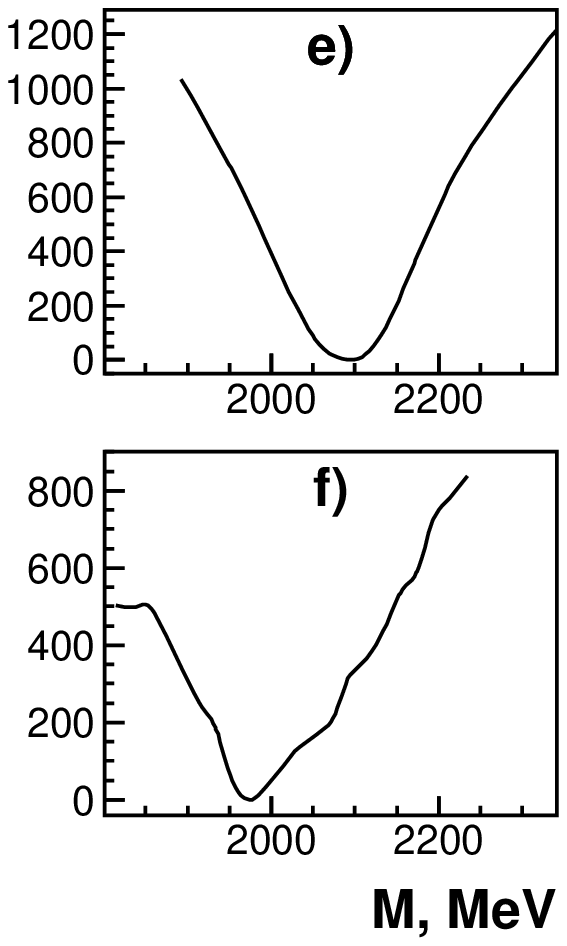,width=0.32\textwidth,clip=on}\\
 \end{tabular}
 \ec
\caption{\label{scan}Mass scan of a relativistic multichannel
Breit-Wigner nucleon resonances. The plots (a) represent the total
likelihood change for $J^P=3/2^+$ and (b) for $J^P=3/2^+$ decaying
into $K\Lambda$ and $K\Sigma$ final states, (c) the total likelihood
change for $J^P=5/2^+$ and (d) for $J^P=5/2^+$ decaying into
$K\Lambda$,  (e) the total likelihood change for $J^P=7/2^+$ for the
solution BG2011-01 and (f) for the solution BG2011-02}
\end{figure*}

In order to visualize the need of this pole, we present in Fig.
\ref{scan}a,b) a scan of the assumed mass. The $J^P=\frac32^+$ wave
is described by a one-pole K-matrix, representing $N_{3/2^+}(1720)$,
plus one Breit-Wigner resonance. The mass of the Breit-Wigner
resonance is changed in discrete steps, all other parameters are
refitted, and the resulting $\chi^2$ is monitored. The fit includes
data on three-body reactions like $\gamma p\to p\pi^0\pi^0$ which
are fitted event-by event using a maximum likelihood method. Twice
the negative likelihood from three-body reactions and the $\chi^2$
from data in binned histograms are added; hence the absolute
$\chi^2$ value has no significance. In Fig. \ref{scan}a) and b), the
$\chi^2$ change $\Delta\chi^2$ is plotted as a function of the
assumed mass. Clear minima are observed in both plots, in
$\Delta\chi^2$ summed over all contributions and when the summation
is restricted to data with $\Lambda K^+$ and $\Sigma K$ in the final
state. The $\chi^2$ changes are similar in magnitude in both plots:
The evidence for $N_{3/2^+}(1900)$ stems from hyperon production
mainly.

{\boldmath $I(J^P)=\frac12(\frac52 ^+)$:\qquad} Real and imaginary
part of the $I(J^P)=\frac12(\frac52 ^+)$ amplitude derived from the
GWU energy-independent analysis \cite{Arndt:2006bf} are shown in
Fig.~\ref{elastic}c. The amplitude houses the well known
$N_{5/2^+}(1680)$. Its properties as derived from our fits are fully
consistent with those given in \cite{Nakamura:2010zzi}. Above this
resonance, a further state has been reported by several groups (see
references in \cite{Nakamura:2010zzi}), but with masses which vary
from 1814\,MeV to 2175\,MeV. In \cite{Hohler:1979yr,Arndt:2006bf},
the evidence for this second $N_{5/2^+}$ state was derived from the
small structure in amplitude (Fig.~\ref{elastic}c) at about 1.9\,GeV
which is present in the energy independent analysis from KH and GWU,
although the structure looks rather different in the two analyses.

Figs.~\ref{scan}c,d show  $\Delta\chi^2$ as a function of the
assumed Breit-Wigner mass in the $I(J^P)=1/2(5/2^+)$ partial wave,
again for the summed $\chi^2$ and for the data with $\Lambda K^+$ in
the final state. Clear minima are observed at about 2050\,MeV; in a
two-pole K-matrix fit the second amplitude pole was found at
$(2050\pm 30)-i(235\pm 20)$\,MeV. The two-pole fit, describing well
the fitted photoproduction data, failed to describe the structure of
the $F_{15}$ elastic amplitude around 1.9\,GeV
(Fig.~\ref{elastic}c); the $\chi^2$ of two-pole fits is about 4.4 -
4.8 for the data in Fig.~\ref{elastic}c. Hence we introduced a
three-pole five-channel K-matrix to describe the elastic amplitude.
This solution reproduces the solution in Fig.~\ref{elastic}c  with
$\chi^2/N_{\rm data}=1.80 - 1.95$.

The three-pole solution, however, shows clear instabilities due to
an over-parameterization of the amplitude. As the result, we find
two solutions: in the first solution, the highest pole is shifted to
even higher masses, to $M=2095^{+30}_{-60}-i\,250\pm 40$; the
position of the second pole is not well defined, it can be located
anywhere in the mass 1800-1950\,MeV region. Its imaginary part
corresponds to a width of $120-300$\,MeV, the main decay mode is the
(open) inelasticity mode assumed to be $N\rho$ (or $N\omega$). By
restricting the inelasticity, a further solution was found. Then,
both poles nearly coincided; they were found at
$(1930\!\pm\!70)-i(200\!\pm\!20)$ and
$(1920\!\pm\!70)-i(270\!\pm\!20)$\,MeV, respectively. If both poles
correspond to real resonances, one of the poles must have large
couplings to $\rho N$ and $\omega N$ while the other couples
significantly to $K\Lambda$. Obviously, further high precision data
are needed to decide if all three poles are required and if so,
which of the solutions is closest to the truth.

In summary, there is certainly at least one resonance above
$N_{5/2^+}(1680)$. It may have a mass in the 2050 - 2100\,MeV mass
range. It is called $N_{5/2^+}(2075)$ in the discussion. However, a
better description of the data is achieved with an additional
resonance in the mass 1800-1950\,MeV region. In the discussion, we
refer to this state as $N_{5/2^+}(1875)$. If this resonance is
introduced, the properties of the third resonance remain uncertain.

{\boldmath $I(J^P)=\frac12(\frac72 ^+)$:\qquad} In this partial wave
one resonance has been reported, $N_{7/2^+}(1990)$. It was observed
by KH \cite{Hohler:1979yr} and CM \cite{Cutkosky:1980rh}, and in an
analysis of data on $\pi N \to \pi \pi N$ \cite{Manley:1992yb}.
Breit-Wigner mass and width were determined to $M=2005\pm 150$\,MeV,
$\Gamma=350\pm100$ in \cite{Hohler:1979yr}, to $M=1970\pm 50$\,MeV,
$\Gamma=350\pm120$ in \cite{Cutkosky:1980rh}, and to $M=2086\pm
28$\,MeV, $\Gamma=535\pm120$ in \cite{Manley:1992yb}. The resonance
is listed in the Review of Particle Properties as $N(1990)F_{17}$.
The resonance was not seen in the GWU analysis \cite{Arndt:2006bf}.
The results of the GWU energy-independent partial wave analysis are
shown in Fig.~\ref{elastic}d,h, together with our fit.

The fit does not converge to one well defined minimum. Instead,
dependent on start values, two classes of solutions are found,
called BG2011-01 and BG2011-02. In the first solution, the pole
position is found at $(1975\pm 15) - i(80\pm15)$\,MeV, in the second
solution at $(2100\pm 15) -i(130\pm 13)$\,MeV. The two solutions
have very different helicity couplings for the $F_{17}$ state;
therefore more precise polarization experiments will be able to
decide which solution is right. A second pole definitely improves
the description but the associated pole is ill-defined, its position
is somewhere between 2300 and 2500\,MeV, the width is large, about
500\,MeV. The main improvements come from the lower-mass pole.

A mass scan in the  $I(J^P)=\frac12(\frac72 ^+)$ wave for the two
solutions is shown in Fig.~\ref{scan}e,f. In the first solution,
$\Lambda K^+$ channels makes a significant contribution to the
signal; in the second solution the mass scan shows significant
minima for $N\pi$ and $N\eta$, and a small dip only for $\Lambda
K^+$.

\begin{table}[pb]
\caption{\label{MassWidth} Pole positions and Breit-Wigner masses of
the positive parity resonances in the region 2 GeV}
\bc
\begin{tabular}{cccccc} \hline\hline
 Sol.&&$N_{1/2^+}(1880)$&$N_{3/2^+}(1900)$&$N_{5/2^+}(2000)$&$N_{7/2^+}(1990)$\\\hline
1.& $M_{\rm pole}$      & $1905\!\pm\!35$   & $1910\!\pm\!12$    & $1850-1950$     &$1975\!\pm\!15$\\
1.& $\Gamma_{\rm pole}$ & $170\!\pm\!40$ &$200\!\pm\!15$& $120-450$       &$160\!\pm\!30$\\
1.& $M_{\rm BM}$        & $1905\!\pm\!40$    &$1918\!\pm\!15$    & $1850-1950$     &$1990\!\pm\!15$\\
1.& $\Gamma_{\rm BW}$   & $210\pm 40$        & $200\!\pm\!20$   & $120-450$       &$160\!\pm\!30$\\
\hline
2.& $M_{\rm pole}$      & $1870\!\pm\!25$   & $1870\!\pm\!35$            & $2050\!\pm\!30$ &$2100\!\pm\!15$\\
2.& $\Gamma_{\rm pole}$ & $280\!\pm\!50$    &$310\!\pm\!40$              & $470\!\pm\!40$  &$260\!\pm\!25$\\
2.& $M_{\rm BW}$        & $1875\!\pm\!30$   &$1890\!\pm\!30$              & $2090\!\pm\!20$ &$2105\!\pm\!15$\\
2.& $\Gamma_{\rm BW}$   & $290\pm 40$       &$315\!\pm\!45$              & $450\!\pm\!40$   &$260\!\pm\!25$\\
\hline\hline
\end{tabular}
\ec
\end{table}

Finally, we mention that the existence of the four positive-parity
nucleons suggested here is obviously fully compatible with the KH
partial wave amplitudes. When the KH amplitudes are replaced by the
GW amplitudes, the full quartet of nucleon resonances is still
required, and their masses and their widths are nearly not affected.

In Table \ref{MassWidth} we summarize our results. For all four
partial waves we find alternative solutions. But in all solutions,
one resonance is required at about 2\,GeV in each of the partial
waves considered here.

In the most straightforward interpretation of the four resonances
$N_{1/2^+}(1880)$, $N_{3/2^+}(1900)$, $N_{5/2^+}(1875)$,
$N_{7/2^+}(1990)$ form a spin quartet of nucleon resonances which
can be assigned to the (D,$L^P_{\,\rm N})=(70,2^+_2)$ multiplet.
Here, $D$ is the SU(6) dimensionality, $L$ is the intrinsic orbital
angular momentum, $P$ the parity and N the shell number in the
harmonic oscillator approximation. This assignment excludes
conventional diquark models: A S-wave diquark is symmetric with
respect to the exchange of the two quarks, a third quark with even
angular momentum is symmetric with respect to the diquark, the
isospin wave function of a nucleon resonance is of mixed-symmetry.
Hence the overall spin-flavor-spatial wave function is of mixed
symmetry. With an antisymmetric color wave function, the overall
wave function has no defined exchange symmetry: the Pauli principle
is violated. The assignment of the four states to a spin quartet,
particularly favored for the solution in which the $N_{7/2^+}(1990)$
mass is 1980\,MeV, rules out the possibility that the {\it missing
resonance problem} is due to a freezing of one pair of quarks into a
quasi-stable S-wave diquark.

Other scenarios are however not excluded: In
\cite{Anisovich:2010wx}, it was suggested that the Pauli principle
could be violated in excited baryons, if the wave function of the
two quarks in the diquark and of the single quark have no overlap.
In this case, the nucleon quartet could be compatible with the
assumption of ``stable" diquarks. Do we have evidence for this
possibility? In the ground states, the Pauli principle is certainly
realized. If the Pauli principle is not imposed for the lowest-mass
negative-parity states, the pattern in the nucleon sector - a
spin-doublet ($N_{1/2^-}(1535)$, $N_{3/2^-}(1520)$) and a spin
triplet ($N_{1/2^-}(1650)$, $N_{3/2^-}(1700)$, $N_{5/2^-}(1675)$ -
should also be observed in the $\Delta$ sector. However, the doublet
($\Delta_{1/2^-}(1620)$, $\Delta_{3/2^-}(1700)$) seems to be too far
separated in mass from the triplet ($\Delta_{1/2^-}(1900)$,
$\Delta_{3/2^-}(1940)$, $\Delta_{5/2^-}(1930)$. Thus the Pauli
principle is required to understand the pattern for $L=1$. When the
Pauli principle is given up for $L=2$, a doublet of positive parity
$\Delta$ states $\frac 32^+$ and $\frac52^+$ is predicted in the
region 1.7-1.8 GeV. Although an observation of the $\Delta_{5/2^+}$
state in the region 1750\,MeV was reported by \cite{Manley:1992yb}
and \cite{Vrana:1999nt}, we did not find any indication for this
state in our present analysis. Hence we conclude that also for
$L=2$, the Pauli principle is applicable.

There is the possibility to assign the four states to other
super-multiplets.  $N_{7/2^+}(1990)$ - with a possible Breit-Wigner
mass of 2105\,MeV from solution BnGa2011-02 - could have $J=7/2;
L=4, S=1/2$ as main intrinsic angular momentum configuration. Its
spin partner could be $N_{9/2^+}(2220)$ with $J=9/2; L=4, S=1/2$.
The two resonances $N_{3/2^+}(1900)$ and $N_{5/2^+}(1875)$ could
form a doublet, too, with $L=2$ and $S=1/2$ coupling to $J=3/2$ and
$5/2$. $N_{1/2^+}(1880)$ could then be a radial excitation with
$L=0$ and $S=1/2$. However, because of their masses, the
interpretation of $N_{1/2^+}(1880)$ as radial excitation and of
$N_{7/2^+}(1990)$ (with mass at 2105\,MeV) as spin-partner of
$N_{9/2^+}(2220)$ seems unlikely, even though not fully excluded.
Thus, the evidence for the quartet of positive-parity nucleon
resonances reported here seems to support symmetric quark model
where all three quarks participate fully in the dynamics.

A third possibility is to organize the four states - together with
negative-parity states reported in \cite{Negpar:2012} - as parity
doublets:
\bc
\begin{tabular}{cccc}
$N_{1/2^-}(1885)$&$ N_{3/2^-}(1860)$&$  N_{5/2^-}(2075)$&$
N_{7/2^-}(2190)$\\
$N_{1/2^+}(1880)$&$ N_{3/2^+}(1900)$&$  N_{5/2^+}(2075)$&$
N_{7/2^+}(2220)$\\
\end{tabular}
\ec
Here, the solution is chosen which yields masses which are best
compatible with the hypothesis that parity doublets exist. The
possibility that baryon resonances are organized in parity doublets
and that parity doublets may reflect restoration of chiral symmetry
in highly excited hadrons \cite{Glozman:1999tk} has attractive
considerably interest \cite{Jaffe:2006jy,Glozman:2007ek}. This is an
exciting possibility which certainly requires further experimental
and analysis efforts.

Summarizing, we have reported results of a partial wave analysis of
a large body of pion and photo-induced reactions. In the forth
resonance region, at about 2\,GeV, at least four positive-parity
nucleon resonances were found, even though some of the solutions
were ambiguous. At the present stage, the data are consistent with
an interpretation of the resonances within symmetric quark models
where three constituent quarks participate in the dynamics (and no
diquark is frozen) and with the conjecture of parity doubling.

We would like to thank the members of SFB/TR16 for continuous
encouragement. We acknowledge support from the Deutsche
Forschungsgemeinschaft (DFG) within the SFB/ TR16 and from the
Forschungszentrum J\"ulich within the FFE program.

\end{document}